\documentclass{aa}  

\usepackage{graphicx}
\usepackage{multirow}
\usepackage{txfonts}
\begin{document}

\title{Atmospheric entry and fragmentation of the small asteroid 2024 BX1: Bolide trajectory, orbit, dynamics, light curve, and spectrum}

   \author{ P.~Spurn\'y\inst{1} \and J.~Borovi\v{c}ka \inst{1} \and  L.~Shrben\'y\inst{1} \and M. Hankey\inst{2} \and R. Neubert\inst{3}
   }

   \institute{Astronomical Institute of the Czech Academy of Sciences, Fri\v{c}ova 298, 25165 Ond\v{r}ejov, Czech Republic \\
              \email{pavel.spurny@asu.cas.cz}    
                    \and
                   American Meteor Society LTD, 54 Westview Crescent, Geneseo, New York 14454, USA
                   \and
            Th\"uringer Landessternwarte Tautenburg, 
Sternwarte 5, D-07778 Tautenburg, Germany
                }

   \date{Received 26 February 2024, Accepted 1 March 2024}
   \titlerunning{Atmospheric entry and fragmentation of the small asteroid 2024 BX1}
   \authorrunning{Spurn\'y et al.}

 
  \abstract{Asteroid 2024 BX1 was the eighth asteroid that was discovered shortly before colliding with the Earth. The associated bolide was recorded by dedicated instruments of the European Fireball Network and the AllSky7 network on 2024 January 21 at 0:32:38–44 UT. We report a comprehensive analysis of this instrumentally observed meteorite fall, which occurred as predicted west of Berlin, Germany. The atmospheric trajectory was quite steep, with an average slope to the Earth’s surface of 75\fdg6. The entry speed was 15.20 km s$^{-1}$. The heliocentric orbit calculated from the 
  bolide data agrees very well with the asteroid data. However, the bolide
  was fainter than expected for a reportedly meter-sized asteroid. The absolute magnitude reached $-$14.4, and the entry mass was estimated to be 140 kg. The
  recorded bolide spectrum was low in iron, based on which, the meteorite was expected to be rich in  enstatite. The recovered meteorites, called Ribbeck, were classified as aubrites. The high albedo of enstatite (E-type) asteroids can explain the size discrepancy. The asteroid was likely smaller than 0.5 meter and should rather be called a meteoroid. During the atmospheric entry, the meteoroid severely fragmented into much smaller pieces already at a height of 55 km under an aerodynamic pressure of 0.12 MPa. The primary fragments then broke up again, most frequently at heights 39--29 km (0.9--2.2 MPa). Numerous small meteorites and up to four stones larger than 100g were expected to land. Within a few days of publication of the location of the strewn field, dozens of meteorites were found in the area we had predicted. }

{}
{}
{}
{}
{}


   \maketitle
\section{Introduction}

On 2024 January 20 at 21:48 UT, Krisztián Sárneczky at the Hungarian Piszkéstető Observatory discovered a small asteroid, 2024 BX1 \citep{MPEC}, which was on a collision course with Earth, as was quickly confirmed by subsequent observations at more than ten other European observatories. As determined by imminent-impact-monitoring services such as the Jet Propulsion Laboratory Scout, the European Space Agency Meerkat, and the Minor Planet Center internal warning system, the collision should have occurred around 0:33 UT in the area of Berlin, Germany \citep{MPEC}. This prediction was subsequently confirmed by the observation of a very bright bolide, which, from locations with clear skies and within range of the bolide, was seen by a large number of mostly casual observers. It was also recorded by instruments, however, both by casual witnesses and by systems specifically designed for the purpose of bolide observations. 

We give a detailed analysis of this bolide using our standard methods as the bolide was recorded by the various instruments of the European Fireball Network and partly by the cameras of the All-Sky7 system. We first describe the instrumental records we used and at which locations they were, then we describe how we determined the atmospheric trajectory and heliocentric orbit of the body, its velocity and deceleration in the atmosphere, its luminosity and fragmentation, the locations at which its debris hit the Earth's surface, and also its composition and physical properties based on the analysis of the spectra we took. Finally, we summarize the main results and compare them to both the precollisional analyses and to the meteorites that were found. Because the meteorites have already been officially named Ribbeck \citep{MetBull}, we call the observed bolide the same hereafter.

 \section{Instruments and data}
\begin{figure}
\centering
\includegraphics[width=\linewidth]{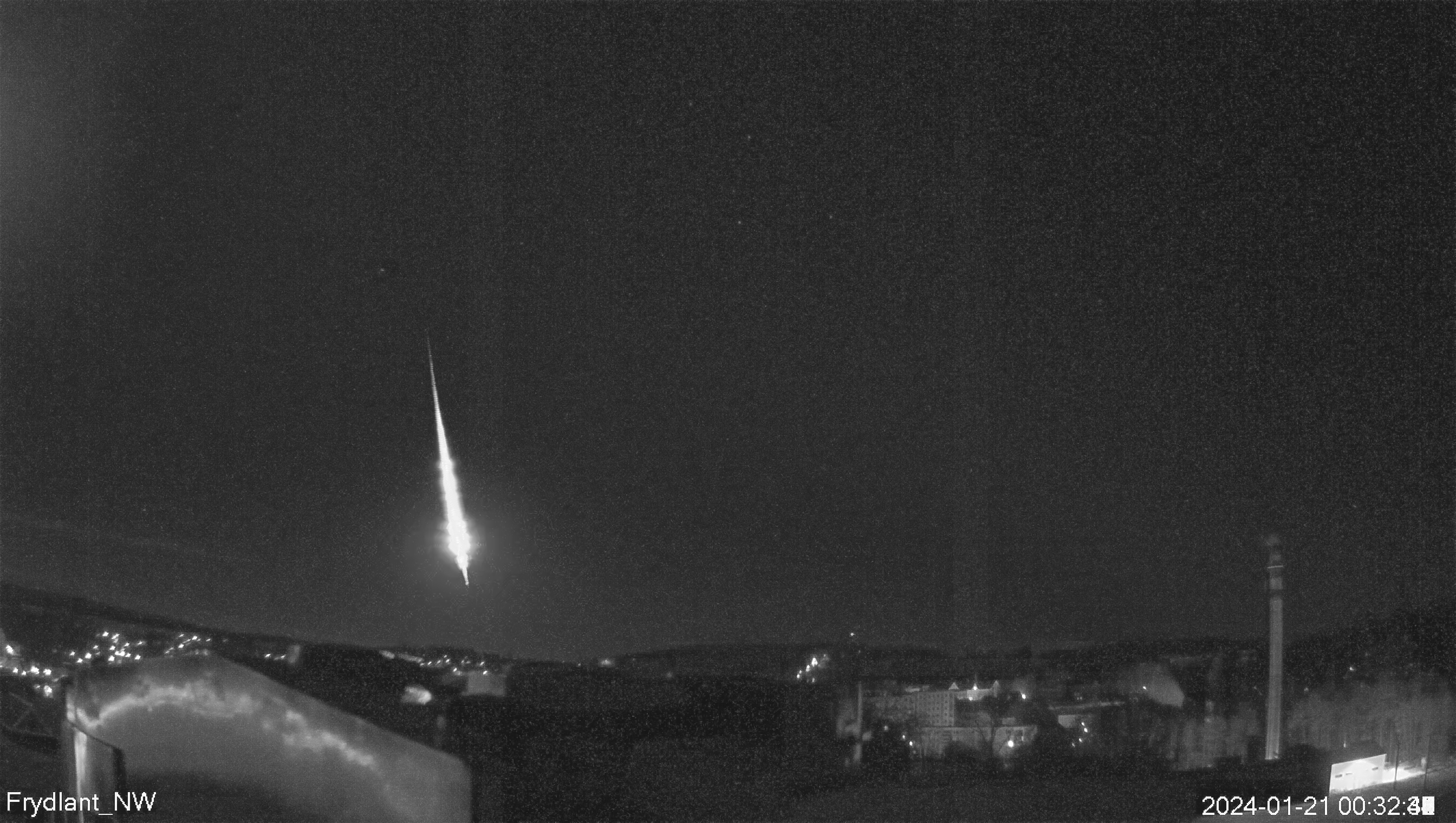}
\caption{Composite image of the Ribbeck bolide from video footage taken by the IP camera at the Frýdlant station.}
\label{bolide}
\end{figure}

\begin{figure}
\centering
\includegraphics[width=\linewidth]{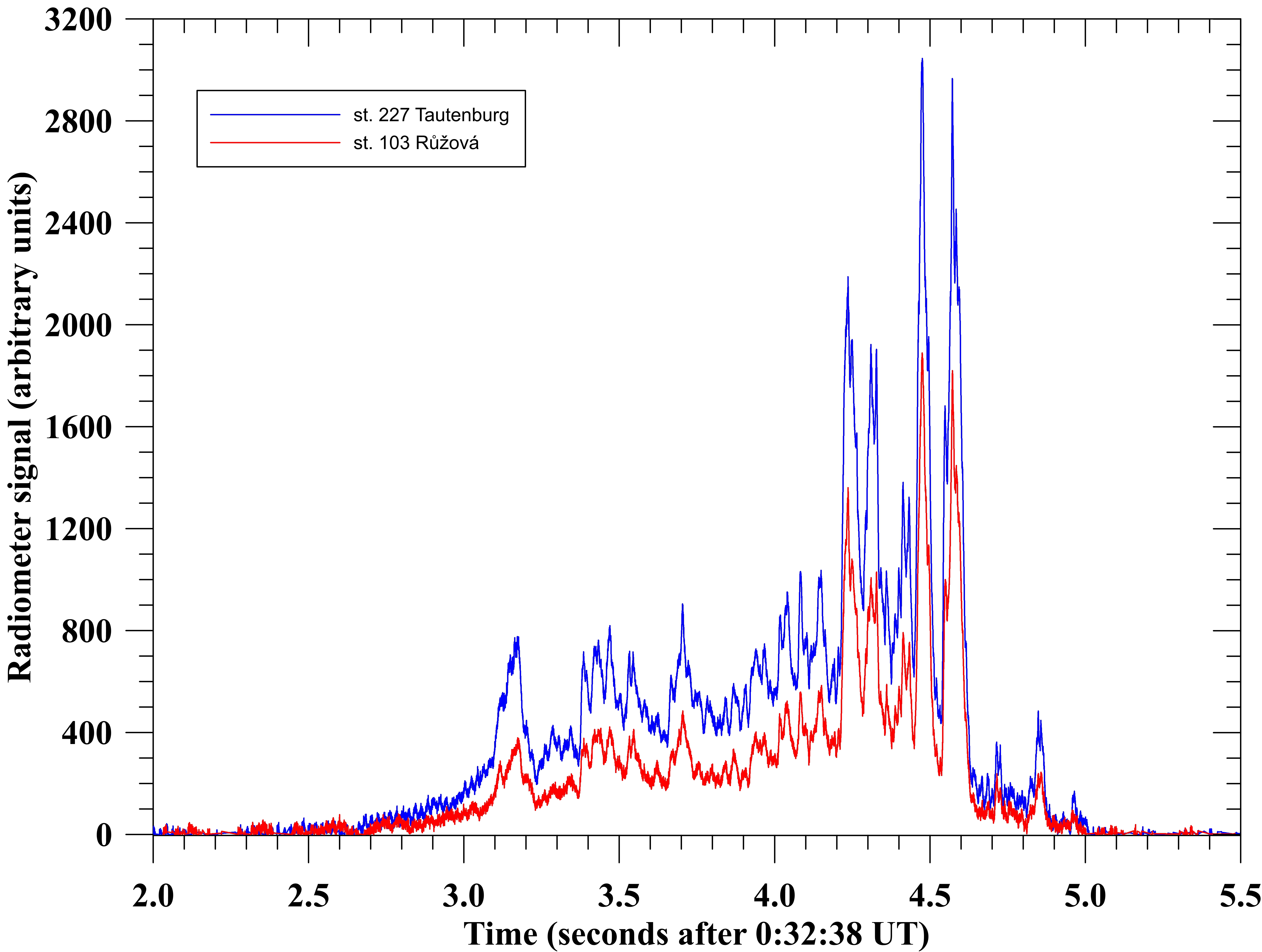}
\caption{Uncalibrated radiometric light curve of the Ribbeck bolide from stations 227 and 103. The reality of every visible detail in the twolight curves is confirmed by independent recordings from two stations 182 km apart.}
\label{lcs2}
\end{figure}

 In this case, the bolide fortunately passed within range of the core of the European Fireball Network (EN), especially the Czech part of the EN, whose center is located at the Ond\v{r}ejov Observatory. Details regarding the current distribution of the stations of this longest-running fireball network in the world and the modern instruments it uses for bolide observations can be found in \cite{Spu17} or \cite{Bor22}. This part of the EN currently consists of 21 stations located in Czechia (15), Slovakia (4), and partly also Germany (1) and Austria (1). The stations are equipped with Digital Autonomous Fireball Observatory (DAFO) which produces all-sky photographic images and radiometric light curves and/or its spectral version (SDAFO). In addition, Internet Protocol (IP) video cameras covering part of the sky are present. For the observation of this event, we were also fortunate in the weather because it was clear over almost all stations. In order to accurately determine the bolide trajectory in the atmosphere and its original orbit in the Solar System (in this case, it was only a confirmation of the already known orbit from pre-atmospheric observations), we used a total of 17 optical records (10 digital all-sky images, and seven video records), 15 of which were from the EN. Two video records were from the German part of the American Meteor Society (AMS) AllSky7 (AS7) network \citep{MH20}. The image of the bolide taken by the video camera from the EN station Frýdlant is shown in Fig.~\ref{bolide}. In addition, we used three radiometric and three exclusive spectral records from the EN network to determine the asteroid properties. In Fig.~\ref{lcs2} we present the uncalibrated radiometric light curve of the bolide taken from Tautenburg Observatory in Germany and the Czech station Růžová as an example of an important record. This was used not only to determine the exact time of the bolide observation (which was determined with an accuracy corresponding to the temporal resolution of the radiometric record, which is 0.2 ms, i.e., 5000 samples per second), but mainly also to determine the exact profile of the bolide luminosity (e.g., the position and amplitude of individual flares), which was used to reveal the fragmentation in the atmosphere (see Section~\ref{fragmentation}). This image also demonstrates the high fidelity of both records, which among other things means the realism of all the details in the displayed light curves. The signal at Tautenburg is higher because this station was closer to the bolide, as shown in Table~\ref{camerastable}, where
 the data both for the position of all stations and for all optical records used are collected. The weights for the trajectory calculations were determined by giving double weight to the digital images compared to the video camera records because these images have a higher resolution. The very close by Ketz\"{u}r station alone was slightly favored due to the short distance from the bolide. The one-third weight for the spectral image (zero order) from Tautenburg was given because only a small part of the trajectory at the edge of the camera field of view was recorded. The velocity was only determined from the video cameras because the bolide was slow and relatively angularly short, and the velocity in the images was therefore not measured reliably. The speed for mainly the first half of the trajectory was best measured on the IP camera in Ondřejov and therefore has twice the weight of the other IP cameras. For the AS7 cameras, the weight was further reduced because much of the bolide path was overexposed and very difficult to measure. In addition to these direct photographic, video, and radiometric records, detailed spectra were obtained. The photographic spectrum was taken with the SDAFO wide-angle camera at Tautenburg, and video spectra were taken with the IP cameras at Ondřejov and Kunžak. The spectral program of the EN was described in more detail in \citet{IMC2018}. All these records were used for the very comprehensive analysis of the bolide that is decribed in the following sections. The raw data from all records used are available upon request. 

\begin{table*}[t]

\caption{Locations of the cameras, their distances to the fireball at the beginning and end, span of the recorded heights, total length, and used weights.}
\label{camerastable}
\begin{tabular}{@{\extracolsep{\fill}} lll|llc|cc|cc|c|c|c}
\hline\noalign{\smallskip}
\multirow{2}{*}{Station} &{Network}& \multirow{2}{*}{Camera} & \multicolumn{3}{c|}{Coordinates (WGS84)} & \multicolumn{2}{c|}{Distance (km)} & \multicolumn{2}{c|}{Height (km)} & Length & \multirow{2} {*} {W$_T$} & \multirow{2}{*} {W$_V$} \\
 &No.&& Longit. E	& Latitude N & h (m) & Beg	& End	& Beg	& End & (km) &&\\
\hline\noalign{\smallskip}
\multirow{2}{*}{Frýdlant}	&\multirow{2}{*}{EN 6}&	DAFO	&\multirow{2}{*}{15.09047}	&	\multirow{2}{*}{50.91773}	& \multirow{2}{*}{350} & 271.5 &	257.5	&	70.33	&	26.50	&	45.3 & 6 & $-$	\\
&&IP&&& & 281.0	&	256.7	&	92.08	&	23.13	&	71.2 & 3 & 4	\\ 
\multirow{2}{*}{Polom}	&\multirow{2}{*}{EN 10}&	DAFO	&\multirow{2}{*}{16.32225}	&	\multirow{2}{*}{50.35015}	& \multirow{2}{*}{748} &	374.5	&	362.6	&	67.49	&	26.16	&	42.7 & 6 & $-$	\\
&&IP&&&& 380.7	&	361.9	&	84.59	&	22.76	&	63.9 & 3 & 4	\\ 
\multirow{2}{*}{Jičín}	&\multirow{2}{*}{EN 8}&	DAFO	&\multirow{2}{*}{15.34047}	&	\multirow{2}{*}{50.43439}	& \multirow{2}{*}{279}	&322.7	&	310.6	&	70.53	&	27.47	&	44.5 & 6 & $-$	\\
&&IP&&&& 330.1	&	309.7	&	90.36	&	23.16	&	69.4 & 3 & 4	\\
\multirow{2}{*}{Kunžak}	&\multirow{2}{*}{EN 2}&	DAFO	&\multirow{2}{*}{15.20093}	&	\multirow{2}{*}{49.10729}	& \multirow{2}{*}{656} &	442.2	&	433.7	&	71.50	&	27.17	&	45.8 & 6 & $-$	\\
&&IP&&& & 447.0	& 433.2 &	89.77 &	23.93 &	68.0 & 3 & 4	\\ 
\multirow{2}{*}{Ondřejov}	&\multirow{2}{*}{EN 20}&	DAFO	&\multirow{2}{*}{14.77994}	&	\multirow{2}{*}{49.91007}	&	\multirow{2}{*}{527} & 349.0 &	340.2	& 69.06	&	29.44	&	40.9 & 6 & $-$	\\
&& IP && && 355.7 & 339.7 & 90.54 & 26.41 & 66.2 & 3 & 8 \\
Růžová & EN 3 &	DAFO & 14.28653	& 50.83411 & 348 & 246.7 & 233.4 & 74.12 & 30.36 & 45.2 & 6 & $-$	\\
Šindelová & EN 1 &	DAFO & 12.59666 & 50.31740 & 595 & 265.0 & 261.5 & 68.95 & 49.28 & 20.3 & 6 & $-$	\\
Přimda & EN 11 & DAFO &	12.67807 & 49.66960	& 752 & 334.1 & 331.2 &	59.03 &	27.20 &	32.9 & 6 & $-$	\\
Kocelovice & EN 5 &	DAFO & 13.83829 & 49.46724 & 525 & 370.1 & 363.9 & 67.74 & 27.12 & 42.0 & 6 & $-$	\\
Tautenburg & EN 27 & SDAFO & 11.71061 &	50.98168 & 338 & 205.5 & 200.3 & 85.61 & 65.38 & 20.9 & 2 & $-$	\\
Ketzür & AMS 16 & AS7 &	12.63124 &	52.49502 & 45 &	82.5 &	26.4 &	80.52\tablefootmark{a} & 21.26 & 61.2 & 4 & 1	\\
Lindenberg & AMS 22 & AS7 &	14.12152 & 52.20867 & 125 & 157.4 & 113.9 & 93.32 & 22.60 & 73.0 & 3 & 1	\\
\hline\noalign{\smallskip}
\end{tabular}
\tablefoottext{a}{entering field of view}
 \tablefoot{(W$_T$) and (W$_V$) are trajectory and velocity weights.}
\end{table*}

 \section{Trajectory and orbit}

 \begin{figure}
    \centering
    \includegraphics[width=1\linewidth]{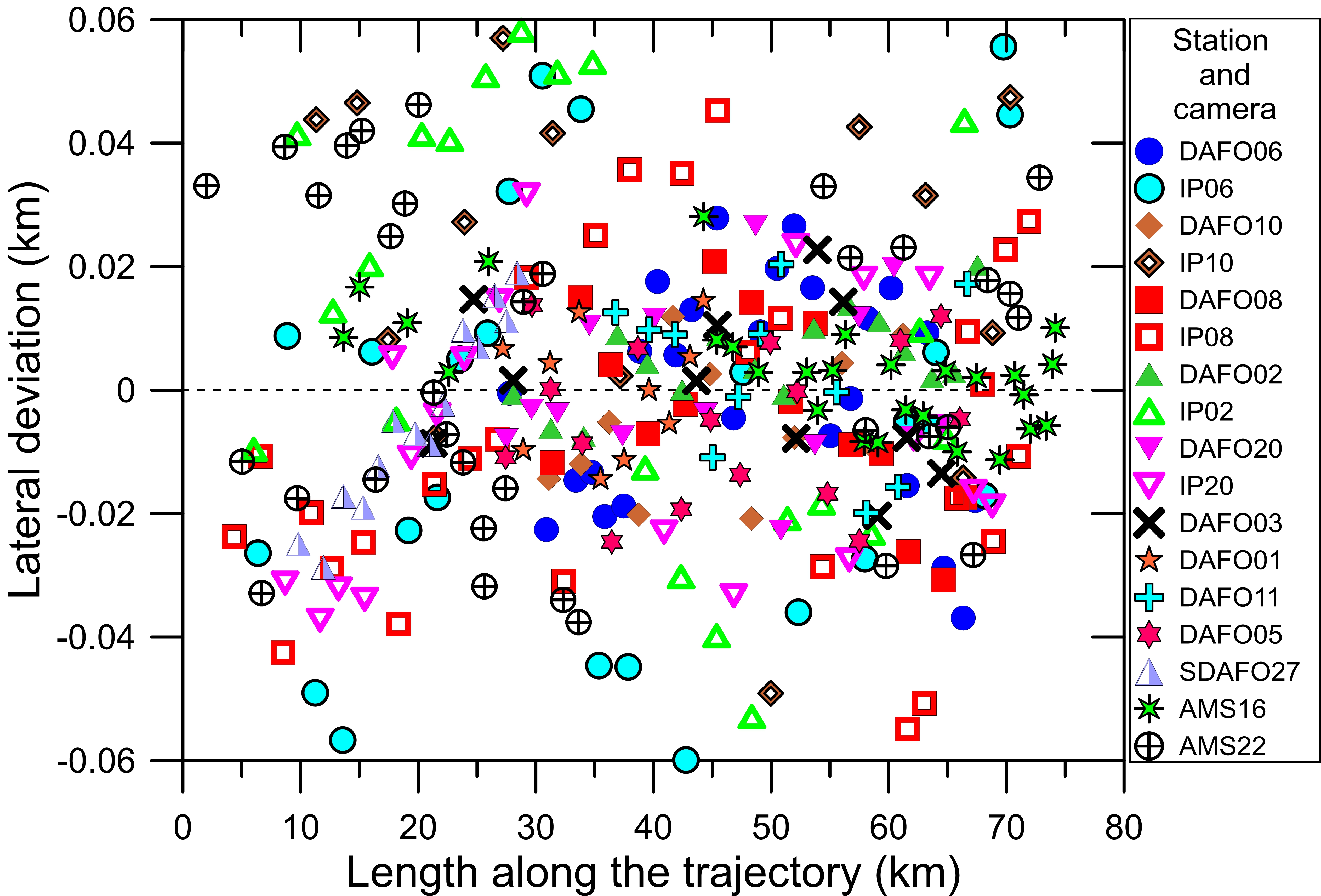}
    \caption{Lateral deviations of all measured points in the luminous path of the fireball from all available records (17). The y-axis scale is highly enlarged, and one standard deviation for any point on the fireball trajectory is only 20 m.}
    \label{DeviationsBX1}
\end{figure}
 
\begin{figure}
    \centering
    \includegraphics[width=1\linewidth]{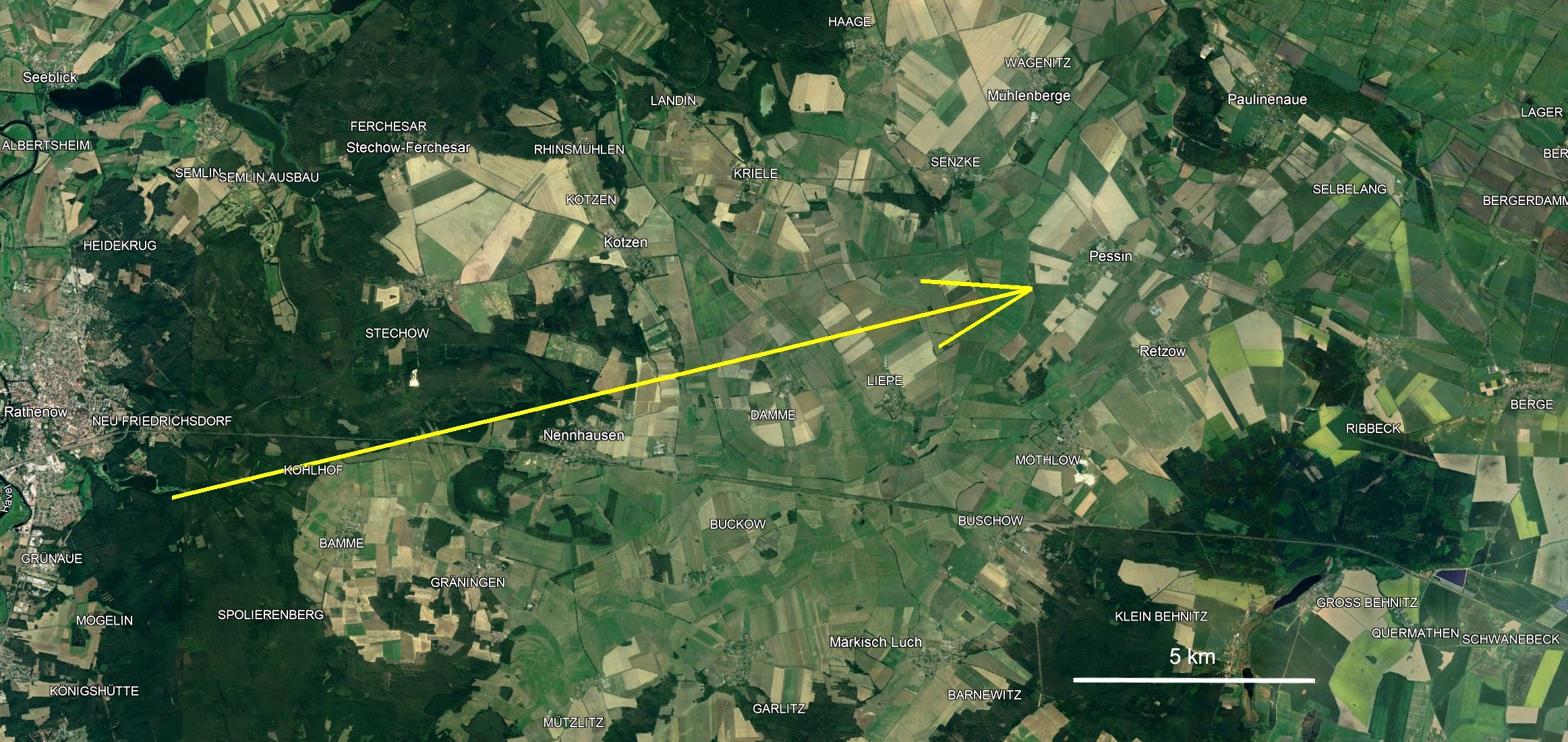}
    \caption{Ground projection of the atmospheric trajectory of the Ribbeck bolide. The average slope was 75\fdg6. The background map is from Google Earth. North is up.}
    \label{atmtraj}
\end{figure}

The bolide trajectory was computed by the least-squares method of \cite{Bor90}. The trajectory was first assumed to be straight. Corrections for curvature due to gravity were applied at the end, when the linear trajectory and velocity were
known. However, due to very steep trajectory, these corrections are very small. The good consistency of all positional measurements is demonstrated in Fig.~\ref{DeviationsBX1}, where the deviations of the lines of sight from the trajectory are plotted. There is no significant systematic trend in any of the 17 records. The points are  randomly mixed. 
All important trajectory data are provided in Table~\ref{trajtable}. The beginning and end points are the points at which the bolide began and ceased to be visible in the records. The exact time corresponding  to the beginning of the bolide shown in the Table~\ref{trajtable} is $0^{\rm h}32^{\rm m}38\fs48$~UT. The ground projection of the atmospheric trajectory is shown in Fig.~\ref{atmtraj} and is relatively short (only 18.5 km), which is due to the steep slope, which was 75.6 degrees on average.  

The geocentric radiant ($\alpha _G, \delta_G$) and heliocentric orbit were computed from the apparent radiant ($\alpha _R, \delta_R$) and entry velocity ($v_\infty$) by the analytical method of \cite{Cep87}, with a small modification accounting for the trajectory curvature described in \citet{Bor22}. 
The entry velocity was determined from the dynamic fit along the first 30 km of the fireball length (at heights above 64 km), where the fireball was measured well and severe fragmentation had not yet begun. The initial mass was set to the value obtained from light-curve modeling (Sect.~\ref{fragmentation}).

The results are given in Table~\ref{orbittable}. In addition, the orbital elements determined from precollision observations are presented in the right column, which provides the unique opportunity to compare the results of the two independent methods. It is obvious and also generally accepted that the orbit determined from observations in interplanetary space is significantly more accurate. However, in this case, the bolide data are also very accurate, and most importantly, all elements agree to within a fraction of one standard deviation determined for the bolide data. The only parameter that differs more than the others is the longitude of the ascending node, which in the method we used \citep{Cep87} corresponds to the length of the Sun at the moment of the collision of the body with the Earth. The problem of this simplification has already been identified by \cite{Clark11} and is also present here. Because the longitude of the ascending node is not a crucial parameter in terms of describing the orbit in the Solar System and does not reduce the accuracy of the orbit determination itself, this small inaccuracy is not very important. The perfect agreement of all other elements, in contrast, is a very important result that fully legitimizes the use of the Ceplecha analytical method.

\begin{table}
\caption{Atmospheric trajectory data of the Ribbeck bolide.}
\label{trajtable}
\begin{tabular}{@{\extracolsep{\fill}} l|r@{ $\pm$ }lr@{ $\pm$ }l}
\hline
& \multicolumn{2}{c}{Beginning} & \multicolumn{2}{c}{Terminal} \\
\hline
Height (km)	&	93.315 & 0.004	&	21.256 & 0.003	\\
Longitude ($^\circ$ E)	&	12.38009 & 0.00008	&	12.64232 & 0.00006	\\
Latitude ($^\circ$ N)	&	52.58959 & 0.00006	&	52.63502 & 0.00005	\\
Slope ($^\circ$)	&	75.557 & 0.008	&	75.74 & 0.02	\\
Azimuth ($^\circ$)	&	74.01 & 0.03	&	74.22 & 0.03	\\
L (km) $/$ T (s) & \multicolumn{4}{c} {74.42$/$ 5.95} \\
\hline
\end{tabular}
 \tablefoot{The height is measured above sea level. The azimuth is taken from the south. L is the length, and T is the duration of the recorded trajectory.}
\end{table}

\begin{table}[t]
\caption{Apparent and geocentric radiant, velocity, and orbital elements (J2000.0) of the Ribbeck meteorite fall.}
\centering
\label{orbittable}
\begin{tabular}{@{\extracolsep{\fill}} r@{ }l|r@{ $\pm$ }lc}
\hline
&&	\multicolumn{2}{c} {Bolide} &	Asteroid\\
\hline
v$_{\infty }$	&	(km/s)	&	15.199 & 0.008&	- \\
$\alpha _R$	&	(deg)	&	119.546 & 0.011 & -	\\
$\delta _R$	&	(deg)	&	46.739 & 0.007&	- \\
$\alpha _G$	&	(deg)	&	114.592 & 0.013 & -	\\
$\delta _G$	&	(deg)	&	44.902 & 0.009&	- \\
v$_G$	&	(km/s)	&	10.476 & 0.012 & - \\
a	&	(A.U.)	&	1.3344 & 0.0007& 1.3343797 \\
e	&			&	0.3740 & 0.0004& 0.3740191	\\
q	&	(A.U.)	&	0.83534 & 0.00016 & 0.8352962	\\
Q	&	(A.U.)	&	1.8334 & 0.0015 & 1.8334632	\\ 
$\omega$ 	&	(deg)	&	243.602 & 0.016 & 243.60428 \\
$\Omega$ 	&	(deg)	&	\multicolumn{2}{c}{300.092} & 300.14142	\\
i	&	(deg)	&	7.264 & 0.009 & 7.26653 \\
P	&	(years)	&	1.5414 & 0.0012 & 1.5414134	\\
\hline
\end{tabular}
 \tablefoot{ Orbital elements from ground-based telescopic observations of the asteroid in space for the epoch 2023 September 13.0 TT are in the right column for direct comparison (MPEC 2024-B76)}
\end{table}
 
 \section{Light curve and fragmentation}
 \label{fragmentation}

 \begin{figure}
\centering
\includegraphics[width=\hsize]{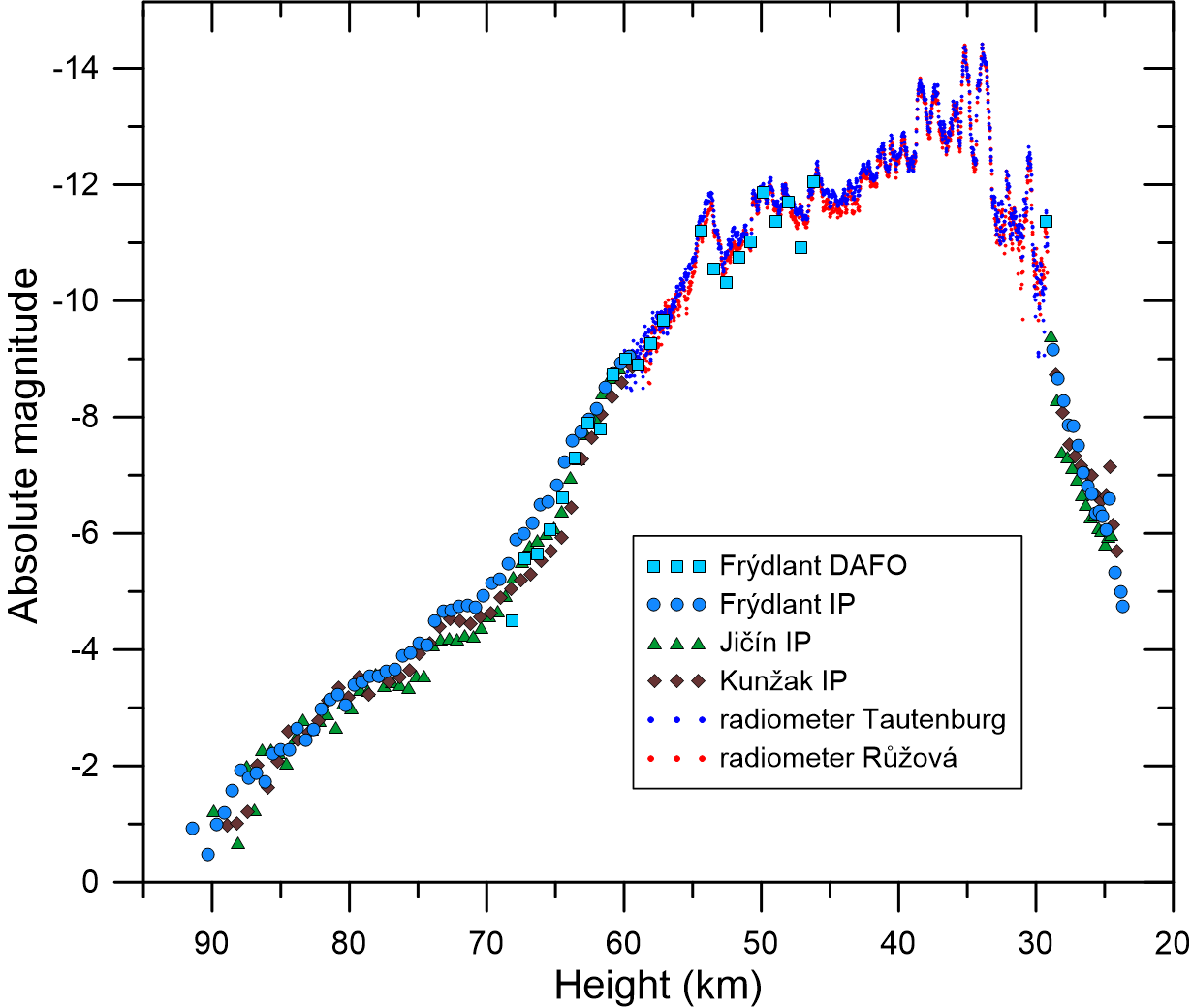}
\caption{Calibrated light curve as a function of height combined from various instruments.}
\label{LC}
\end{figure}

The calibrated bolide light curve as a function of height is presented in Fig.~\ref{LC}. The light curve at the beginning and end
was measured using three IP video cameras. These cameras became strongly saturated during the bright phase of the bolide. The 
medium-brightness part was measured by the photographic DAFO camera in Fr\'ydlant. The absolute calibration (using stars) 
from all these systems agrees well. The bright part of the bolide was well covered by the radiometers, which are
linear detectors with a high dynamic range. The absolute magnitude scale was adjusted to match the camera data.

The bolide reached a maximum brightness of $-14.4$ absolute magnitude in two almost equally bright flares
at heights of 35.2 and 33.9 km. Many other flares of various amplitudes occurred between 54 -- 29 km.  
Another last observed flare occurred at 24.5 km.

\begin{figure}
\centering
\includegraphics[width=\hsize]{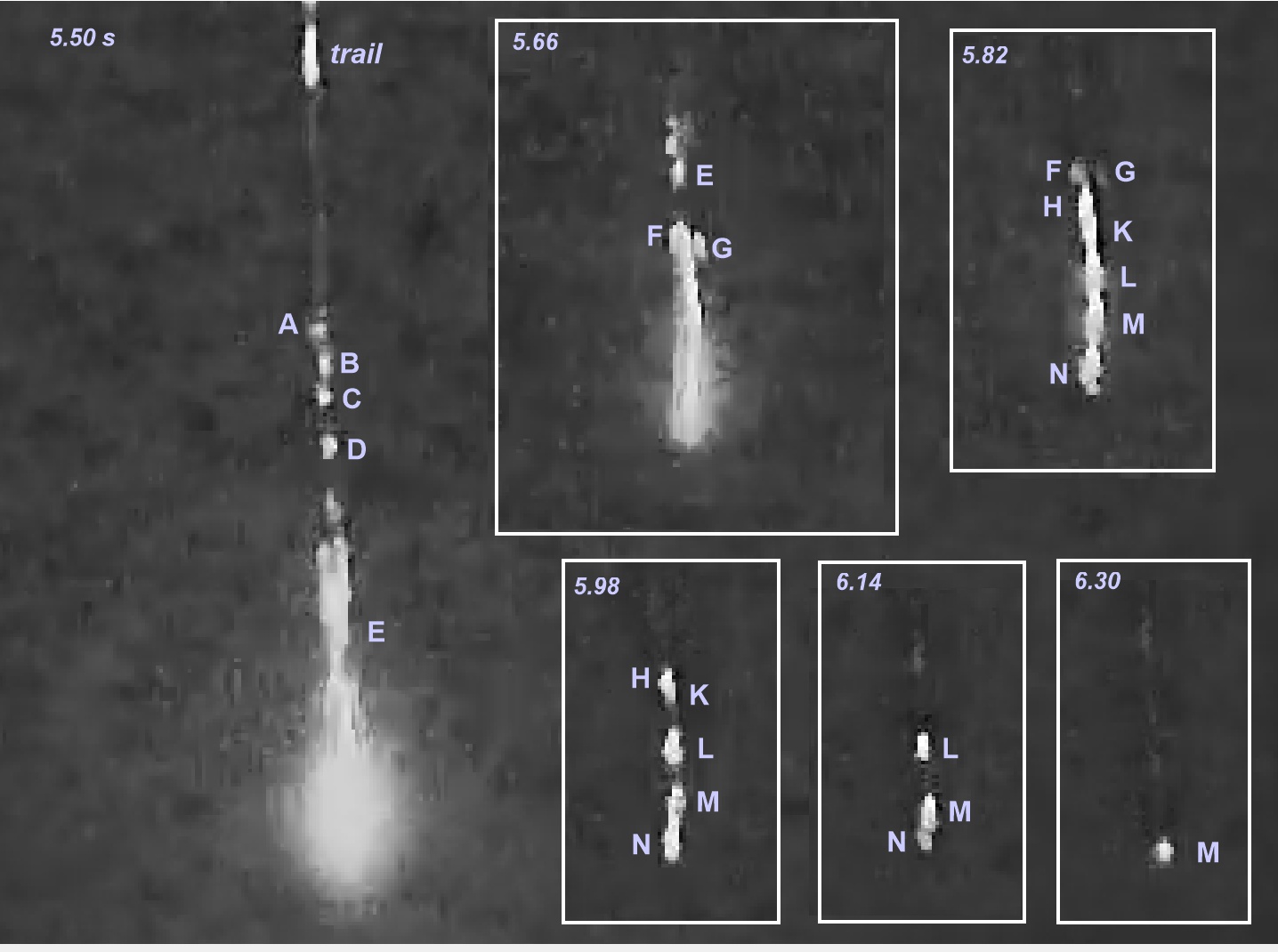}
\caption{Bolide images in six individual frames from the ALLSKY7 AMS 16 video. Fragments that could be measured are marked with capital letters.
The time is given in seconds from 0:32:38 UT.}
\label{fragmentimage}
\end{figure}

The flares are signs of meteoroid fragmentations, at which fine dust was released and vaporized quickly. 
Numerous macroscopic fragments are directly visible on video records taken from a close distance (Fig.~\ref{fragmentimage}). 

We modeled the fragmentation using the semiempirical fragmentation model described in \citet{AJ2020}.
The inputs were the light curve and the dynamics (deceleration) of the bolide as a whole and
individual fragments. Since the spectrum was poor in iron (see Sect.~\ref{spectrum}) and because iron contributes many spectral lines, 
the assumed luminous efficiency (at 15 km s$^{-1}$) was decreased to 4\% for the high-mass limit and 2\% for fine dust
from the 5\% and 2.5\% commonly used, respectively. The product of the drag and shape coefficient was set to 
$\Gamma A = 0.7$, the meteoroid density was assumed to be $\delta = 3100$ kg m$^{-3}$, and the ablation
coefficient was fixed at $\sigma = 0.005$ kg MJ$^{-1}$. The atmospheric density model CIRA72 was used.

\begin{figure}
\centering
\includegraphics[width=\hsize]{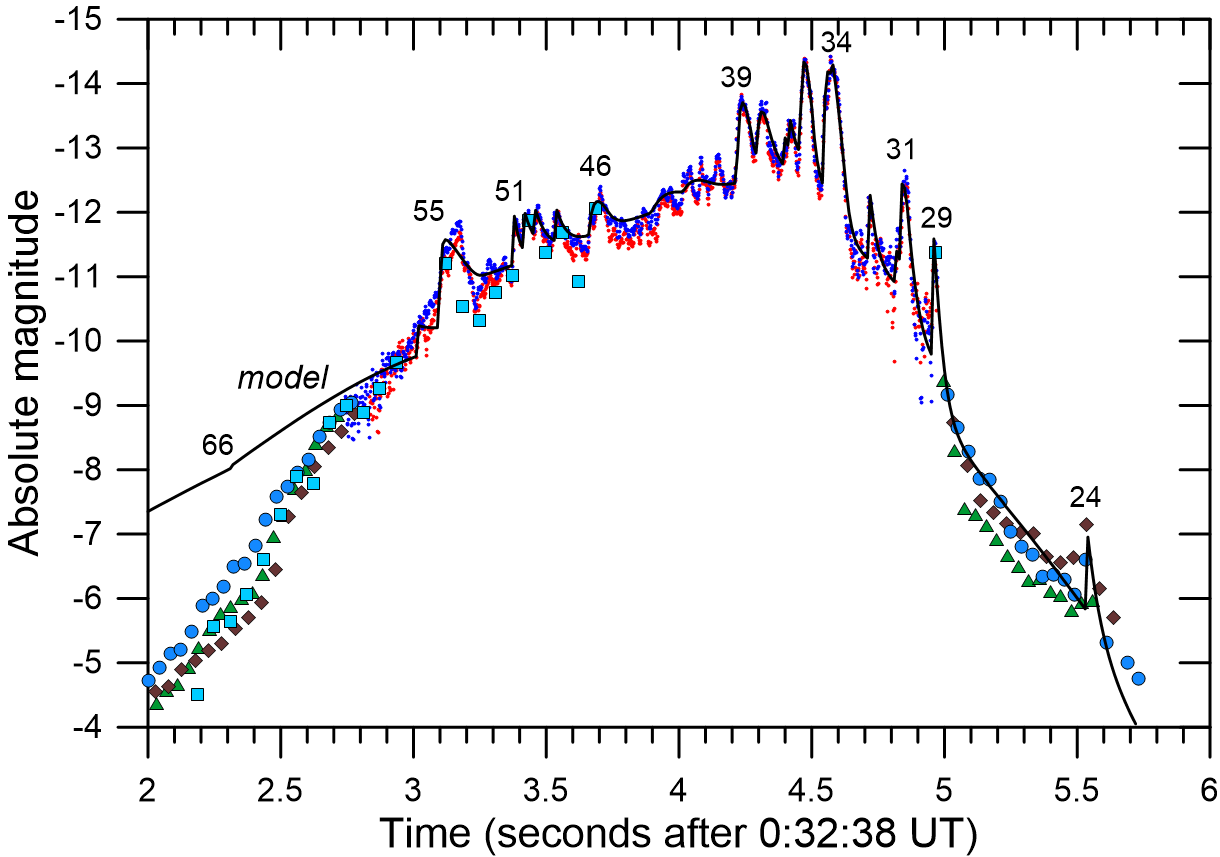}
\caption{Observed (symbols) and modeled (solid line) light curve as a function of time. The symbols are the same as in Fig.~\ref{LC}.
The numbers give the approximate heights in kilometers of selected light-curve features.}
\label{LC_model}
\end{figure}

\begin{figure}
\centering
\includegraphics[width=\hsize]{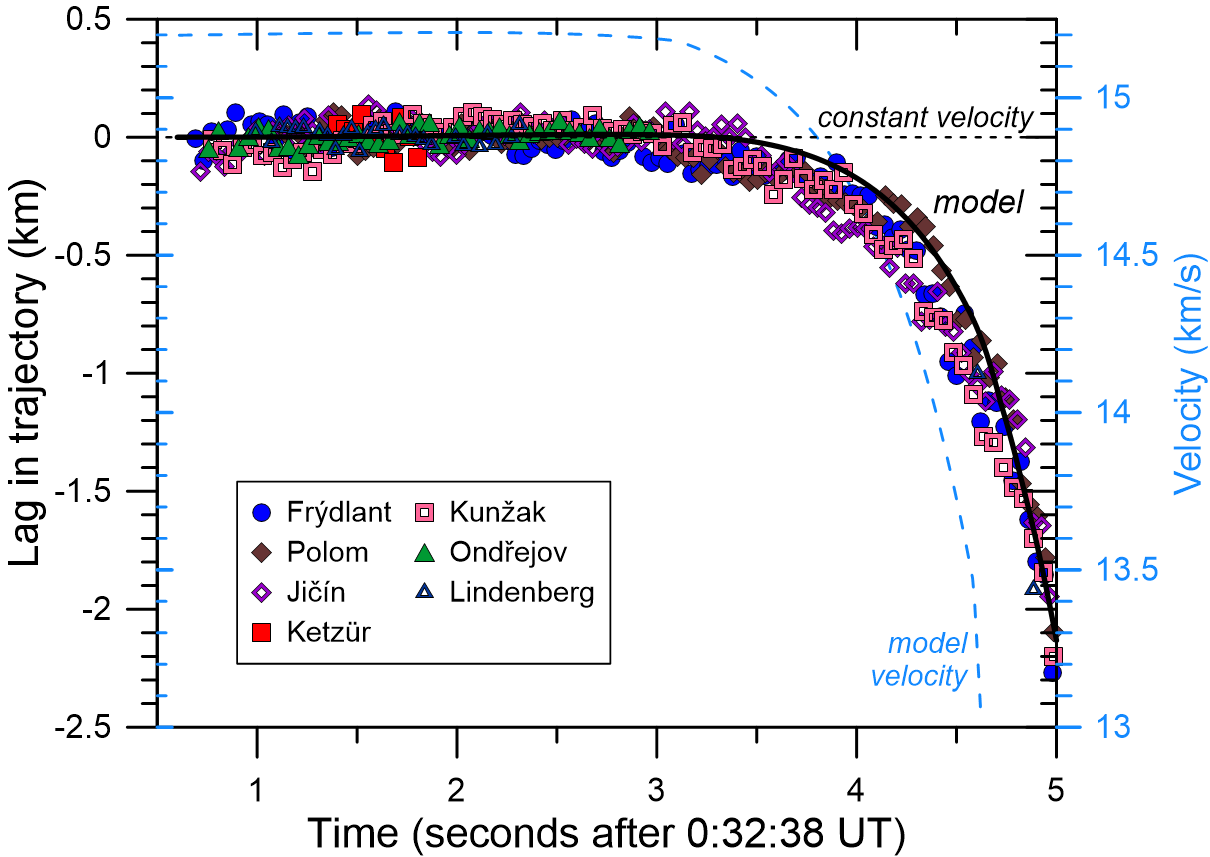}
\caption{Lag in trajectory with respect to a constant velocity of 15.199 km s$^{-1}$ as a function of time.
The symbols represent measurements on individual videos. The solid line is the lag model for the leading fragment. The dashed line
(with the scale on the right) is the corresponding model velocity.}
\label{deceleration}
\end{figure}

The modeled light curve is presented in Fig.~\ref{LC_model}. To fit the entire produced radiation, the initial meteoroid mass was set to 140 kg.
The beginning of the bolide, when the meteoroid was heating up and the 
ablation was not yet in its full stage, cannot be described by the model. It seems that some fragmentation, more precisely, gradual dust loss,
started already during this phase, at a height of 66 km. This is evidenced by a change in the slope of the light curve, but also by the long wake
of the bolide that is visible in the videos at heights of 64 -- 57.5 km. The first major, and in fact catastrophic, fragmentation started at a height of 55 km
and was demonstrated by the first flare. In addition, a significant deceleration of the bolide started after this
fragmentation, as shown in Fig.~\ref{deceleration}.  The deceleration corresponds to a mass of largest fragments of about 5 kg.
At the times 4 -- 4.5~s, the measured lag behind the constant velocity course was even longer than modeled
on most videos (Fig.~\ref{deceleration}). This suggests that the majority of fragments were even smaller (1--2 kg) and that the photocenter was shifted behind
the leading fragments.

The fragments produced at 55 km fragmented further at lower heights. As evidenced by the flares, some minor fragmentations occurred at
heights of 51--40 km, and the major fragmentation occurred at heights of 39--29 km. Dust was released either immediately or quickly because the flares have a high
amplitude and short duration. To estimate the masses and heights at which the surviving fragmentation products originated, the dynamics of the fragments seen
in the Ketz\"{u}r video was measured and fit. We also measured the fragments in the video taken by Michael Aye (at allplanets) in Berlin using a handheld
mobile phone and published on the Internet\footnote{https://mastodon.online/@michaelaye/111791284304249289}.
An attempt to reconstruct the individual fragment trajectories in space failed because the measured trajectories are short and
the data precision is not high enough for this task. All fragments were therefore assumed to follow the same trajectory. 

\begin{table}
\caption{List of the most important fragmentations.}
\label{fragmentationtable}
\begin{tabular}{lllll}
\hline
Time & Height& Aerodynamic& Parent & Product \\
& & pressure & mass & mass \\
 (s) & (km) & (MPa) & (kg)&(kg) \\
 \hline
 2.32 & 66 & 0.03 & 140 & 138 \\
 3.02 & 55.8 &0.11 & 138 & 115 \\
 3.10 & 54.6 & 0.12 & 115 & 5.5 \\
 4.22 & 38.5 & 0.87 & 5.1 & 2.2 \\
 4.30 & 37.7 & 1.07 & 4.4 & 1.5 \\
 4.42 & 35.8 & 1.32 & 3.9 & 0 \\
 4.45 & 35.4 & 1.40 &  4.6 & 1.4 \\
 4.58 & 33.7 & 1.74 & 4.9 & 0.13 \\
 4.72 & 32.0 & 1.93 & 1.3 & 0.5 \\
 4.84 & 30.6 & 2.17 & 1.89 & 0.99 \\
 4.96 & 29.3 & 2.04 & 0.44 & 0.04  \\
 5.54 & 24.4 & 1.79 & 0.78 & 0.07 \\
  \hline
  \end{tabular}
  \tablefoot{The mass of the
body before fragmentation and the mass of the largest fragmentation product in our model
are given in the last two columns.}
 \end{table}
 
 Table~\ref{fragmentationtable} lists the most important fragmentation events. We list the time (counted from 0:32:38 UT),
 height, aerodynamic pressure computed as $p=\rho v^2$, where $\rho$ is the atmospheric density and $v$ is the velocity, 
 the mass of the fragmenting body and the mass of the largest fragmentation product. The fragmentation
 sequence could not be determined unambiguously. Moreover, fragmentations of different fragments likely occurred nearly simultaneously.
 The table lists the fragmentations of the largest fragments in our model. The masses of the fragmentation products were partly
 derived from the dynamics of directly visible fragments (Fig.~\ref{fragmentimage}). Their parameters are given in Table~\ref{fragmenttable}.
 The earliest visible fragments were produced at heights of around 40 km  and had masses in the range $\approx$ 15 -- 25 g.  
 They may have produced meteorites in the mass range of 10 g. The largest piece that emerged from multiple fragmentations at heights 35 -- 29 km (marked N1)
 initially had a mass of almost 1 kg. However, it lost most its mass in another breakup at 24.4 km, and a meteorite of only about 60 g remained.
 The late breakup occurred when the aerodynamic pressure was already decreasing (the velocity at the breakup was only 6.6 km s$^{-1}$).
 The largest fragment (M) that survived at least until the end of the luminous phase of the bolide had a mass of about 400 g.

 \begin{table}
\caption{Individually observed fragments (see Fig.~\ref{fragmentimage}).}
\label{fragmenttable}
\begin{tabular}{llll}
\hline
Designation & Origin& Initial & Meteorite \\
& height & mass & mass \\
  & (km) & (g)&(g) \\
 \hline
A& 43.0 & 16 & 9 \\
B& 41.4 & 19& 11 \\
C& 38.5 & 14& 8 \\
D& 38.5 & 26& 15 \\
E& 32.0 & 40& 25 \\
F& 33.7 & 130& 85 \\
G& 29.3 & 40& 30 \\
H& 35.3 & 270& 170 \\
K& 32.0 & 180& 120 \\
L&  32.0 & 300& 210 \\
M& 35.4 & 660& 410 \\
N1\tablefootmark{a}& 30.6 & 990& -- \\
N& 24.4 & 70& 60 \\
  \hline
  \end{tabular} \\
  \tablefoottext{a}{N1 is the precursor of N.}
 \end{table}

 \section{Meteorite strewn field}

  \begin{figure}
\centering
\includegraphics[width=\hsize]{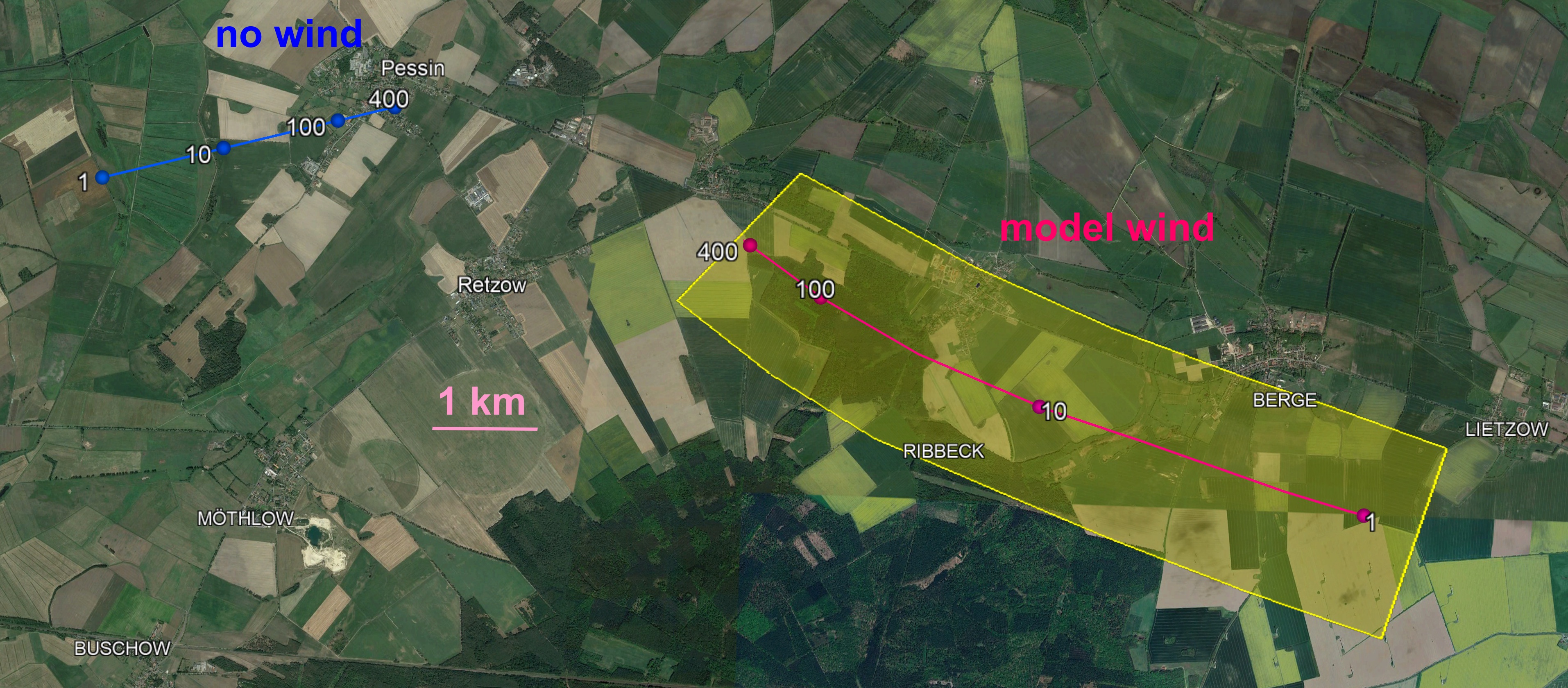}
\caption{Map of the computed meteorite fall locations for the hypothetical case of no atmospheric wind (blue)
and using the ALADIN wind model (red). The numbers show the meteorite masses in grams 
and are valid for spherical meteorite shapes. The estimated uncertainty area is shown in yellow for the wind case.
The 400 g point lies at 52\fdg6285 N, 12\fdg7200 E. The background map is from Google Earth. North is up.}
\label{strewnmap}
\end{figure}

The expected meteorite landing positions were computed using the dark-flight procedure of \citet{Cep87}.
The procedure was started when the fragment velocity dropped to 2 km s$^{-1}$. At this time, the ablation surely ceased. 
The largest fragment M was visible in the video (and thus ablating) at a velocity of 2.7 km s$^{-1}$.
The atmospheric pressure, temperature, and wind profile for 0 UT and 1 UT was kindly provided to us by 
R. Bro\v{z}kov\'a from the Czech Hydrometeorogical Institute using the Aire Limit\'ee Adaptation dynamique D\'eveloppement InterNational  (ALADIN) weather forecast model.

A map of the computed strewn field is shown in Fig.~\ref{strewnmap}. For comparison, the computation without wind
is shown as well. In this case, the meteorites would be distributed along a line in the direction of the bolide flight, with
large meteorites farther ahead. The wind, blowing from west to northwest directions, completely changed
the situation. The meteorites were distributed along the wind direction, with the smallest meteorites blown farther away.
The distance flown depends not only on the meteorite mass, but also on the shape, which affects the drag coefficient
during the dark flight \citep{Towner}. The masses given in Fig.~\ref{strewnmap} are for spherical shapes. Brick-like
or irregularly shaped meteorites may have been transported farther east.

Fig.~\ref{strewnmap} also shows the uncertainty in the meteorite line, which is mostly due to the uncertainty in the wind. The wind
can exhibit variations in time that the wind model cannot account for. Some spread of meteorites can be also expected 
due to ejection angles during the fragmentations and due to aerodynamic effects. Some of the meteorites can therefore lie outside
the marked area.

The number of meteorites is uncertain. Four meteorites were probably larger than 100 g at the end of ablation
(Table~\ref{fragmenttable}). Nevertheless, further breakups during the dark flight cannot be excluded.
Smaller meteorites were surely more numerous than given in Table~\ref{fragmenttable} because there are unresolved
fragments in the video, for example, between fragments E and D (Fig.~\ref{fragmentimage}). There may be several dozen meteorites 
with masses of 10 -- 100 g and up to several hundred of those between 1 -- 10 g.

Our map in Fig.~\ref{strewnmap} was made available already on January 22, and dozens of meteorites were already found
in the designated area \citep{MetBull}. The exact locations of most of them are not known to us at the time of writing, however.
 
\section{Spectrum and composition}
\label{spectrum}

 \begin{figure}
\centering
\includegraphics[width=\hsize]{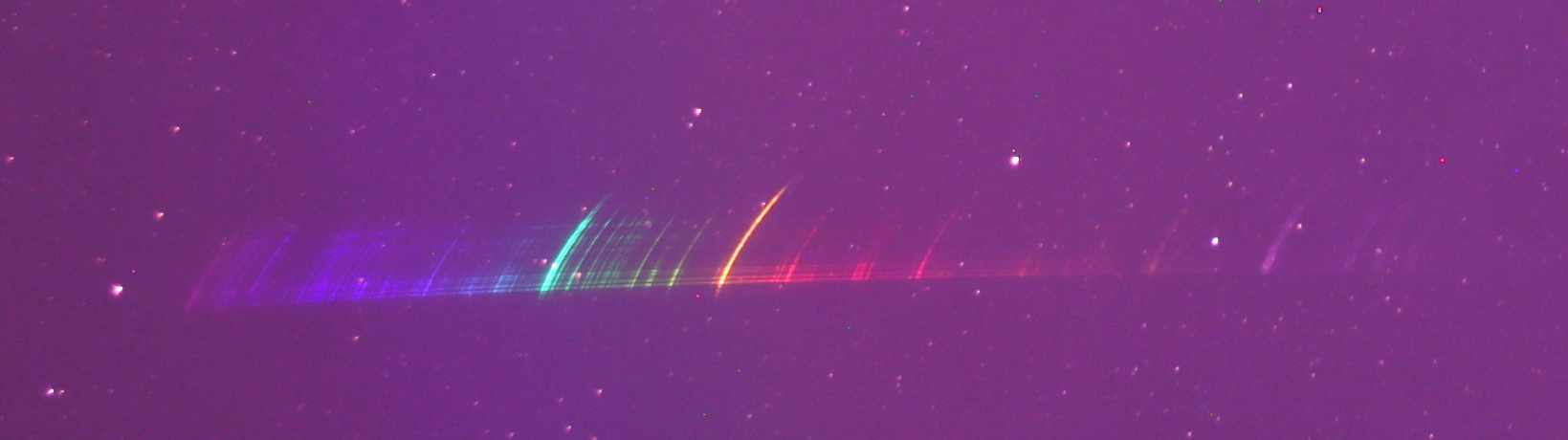}
\caption{Original color image of the Ribbeck bolide spectrum with stars in the background. The bolide moved from top to bottom.
The wavelengths increase from left to right. The curvature of the spectral lines is due to the geometry of the fish-eye lens.}
\label{spectrumimage}
\end{figure}
 
\begin{figure*}
\centering
\includegraphics[width=0.9\hsize]{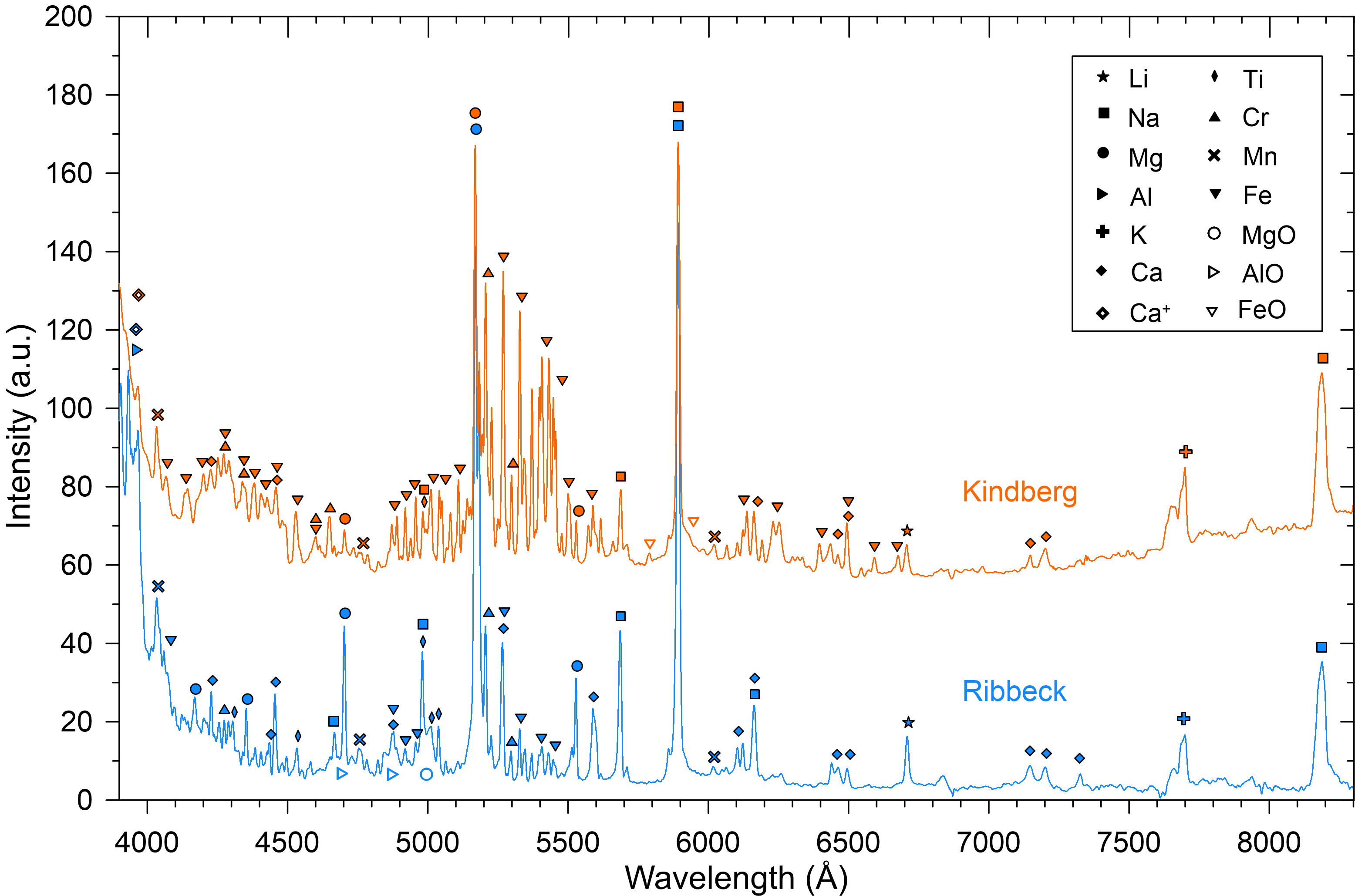}
\caption{Comparison of spectra of the Kindberg and Ribbeck (2024 BX1) bolides in the range 3900 -- 8300~\AA.
The Kindberg spectrum has been vertically offset by 50 units. Atoms, ions, and molecules contributing to the observed lines and bands
are identified using the symbols from the legend. In some cases, two atoms contribute to one line. In other cases, one symbol
is used for two or more neighboring lines of the same atom(s).}
\label{plotspectra}
\end{figure*}

The bolide spectrum taken by the SDAFO at Tautenburg is shown in Fig.~\ref{spectrumimage}. The bolide was low over the horizon and angularly
short, but bright enough to produce a good spectral image. The bright green line belongs to magnesium, and the yellow line is that of sodium.

The spectrum is best inspected in comparison with another spectrum.  In Fig.~\ref{plotspectra}, the spectrum of the
bright part of the bolide (heights 34--39 km) is plotted together with the spectrum of the Kindberg bolide at similar heights (38--40 km).
The Kindberg bolide led to the fall of an ordinary chondrite of type L6 in Austria on November 19, 2020 \citep{Kindberg}. The spectrum was taken
by another SDAFO of the EN network at Chur\'a\v{n}ov station. The two bolides had similar speeds.

Although the same Mg and Na lines are the brightest lines in both spectra, there is a significant difference in that the Kindberg
spectrum is otherwise dominated by numerous Fe lines, the brightest of which lies between 5250--5500~\AA. In Ribbeck,
the Fe lines are much fainter or absent, and many lines of Mg, Na, and Ca are visible. The lines of K, Li, Mn, Cr, and Ti are visible as well.
All these lines are present in Kindberg as well, but they are less conspicuous. The bands of AlO and MgO oxides, which were previously only observed in the
very bright Bene\v{s}ov bolide \citep{Benesov-molecules}, are also visible, while FeO is present in Kindberg, but not in Ribbeck.
The lines of Al seem to be bright in Ribbeck, but are blended with Ca$^+$ and difficult to measure.

A rough estimate of the abundances of neutral atoms in the radiating plasma indicates that Fe was depleted 30--50 times
relative to Mg in Ribbeck in comparison with Kindberg. Li, Ca, and Ti have nearly the same ratio to Mg. K may be twice
underabundant, and Mn is two to three times and Cr is eight times underabundant. Na seems to be two to three times enhanced in Ribbeck.
A more rigorous abundance determination will be conducted in the future. Nevertheless, a strong depletion of Fe is evident and indicated 
an enstatite-rich (MgSiO$_3$) material even before the meteorites were found.

\section{Discussion and conclusions}

The combination of the precollision observation of an asteroid, the associated bolide, and the recovery of meteorites is the fourth such case in history. However, 2024 BX1 is clearly the best-documented case of this type. The first such case was the observation of asteroid 2008 TC3, which was associated with the Almahata Sitta meteorite fall in Sudan on 7 October 2008
\citep{Jen09}. The second similar case was associated with the fall of asteroid 2018 LA in Botswana on 2 June 2018 \citep{Jen21}. However, for both of these falls, the data for the bolide were limited. The third such case was the fall of asteroid 2023 CX1 in France on 13 February 2023 \citep{VidaACM}. For this case, the documentation of the bolide is much better, but the data from France are mostly amateur records, even though it was within the FRIPON network\citep{FRIPON}, which did not observe the bolide properly, however. Fortunately, the 2024 BX1 bolide occurred within the range of the EN network, which provided precise and multi-instrumental data, including a detailed radiometric light curve and spectrum. The video data from the AllSky7 network, taken from a closer distance,
were also useful, especially for measuring the fragments, and proved to be reliable when properly reduced.

 In two main areas can asteroid and bolide data be directly compared: the trajectory in the atmosphere, and especially, in the original orbit in the Solar System. The atmospheric trajectory of the body will clearly be more accurately determined from observations of its passage through the atmosphere, that is, from direct data from the bolide networks. This is summarized in Section 3 in Table~\ref{trajtable} and Fig.~\ref{DeviationsBX1}, and the position of the trajectory in the atmosphere is  determined to an accuracy of about 20 meters, which can hardly be reached by precollisional observations. On the other hand, it is generally accepted that the determination of the orbit in the Solar System mainly is the domain of observations from telescopes when the body is in interplanetary space. This is practically the first time that we can compare the two methods independently. This comparison is presented in Table~\ref{orbittable} and shows that both methods give very similar results, that is, all elements determined based on the bolide observations fit within one standard deviation (for most of elements, it is only small fraction of one standard deviation) to the elements determined before collision. These are clearly more precise. Therefore, it is evident that the bolide data, if properly and thoroughly processed, can give a perfectly plausible description of the orbit in the Solar System. This is one of the absolutely crucial results, and it moreover substantially supports the plausibility of most of the other conclusions from the bolide analysis as well.

We can also compare the size determination.
Asteroid 2024 BX1 was observed in space and had an absolute magnitude 32.84 \citep{MPEC}.
Based on this, the size was estimated and widely reported before the impact as about one meter. 
The corresponding asteroid mass would be about 1700 kg. 
The estimate was based of the usual albedo of the most common S-type asteroids of 0.15.
The observed bolide, though surely very nice,  
was nevertheless much fainter than could be expected for a mass like this. Our initial mass estimate based on the bolide radiation
was only about 100 kg. After inspecting the spectrum, we found that the material was significantly poorer
in iron than typical meteoroids. A composition rich in enstatite was suggested. The high albedo of about 0.50 of enstatite-rich
E-type asteroids and aubrite meteorites \citep{Clark, Dibb} would change the asteroid size estimate to 0.50 m.
Subsequently, the meteorites were recovered and indeed classified as aubrites \citep{MetBull}.
This was the first time that the meteorite type was predicted from the bolide spectrum.
Our final estimate of the asteroid mass from the bolide model is 140 kg. The corresponding size is 0.44 m, if we use the typical aubrite density
3100 kg m$^{-3}$ \citep{Britt}. Considering the uncertainties in the actual luminous efficiency, albedo, and bulk density, the
agreement between bolide and asteroid data is good.

A simple method for estimating the meteoroid diameter from the bolide
absolute magnitude and velocity at a height of 60 km was recently proposed by \citet{Johnston}. From their formula and the Ribbeck magnitude of $-9$ at 60 km, we obtained a diameter of 34 cm. This is too small, in our opinion, and the reason may again be the lack of iron. Except for sodium, the iron lines are usually among the brightest in the spectra of early bolide parts, while Mg and especially Ca appear later on. Since the iron lines were faint in Ribbeck, it was fainter at 60 km than an ordinary bolide. A magnitude $-9.7$ would be expected for a diameter of 44 cm according to the formula of \citet{Johnston}.

Because it was smaller than one meter, 2024 BX1 should rather be referred to as
a meteoroid than an asteroid, according to the adopted definitions of the
International Astronomical Union commission F1 \citep{definitions}. Regardless of
the term, 2024 BX1 was probably the smallest natural body ever observed telescopically
in space.

The atmospheric entry was characterized by extensive fragmentation. Ordinary chondrites usually fragment in two distinct
phases, at 0.04--0.12 MPa and 0.9--5 MPa \citep{AJ2020}. The behavior of 2024 BX1 was similar, with two differences.
First, numerous fragmentations, though minor, also occurred between the two phases. Second, the first phase was 
extraordinary severe. The masses of the largest surviving fragments were lower than 5\% of the original mass. Of
the bodies studied by \citet{AJ2020}, only the Bene\v{s}ov meteoroid fragmented this severely at the beginning. In this case, this was not very surprising because
Bene\v{s}ov was a conglomerate of different meteorite types \citep{Benesov}. It seems that 2024 BX1 was a weekly cemented conglomerate of small boulders, that is, type C material as defined by \citet{AJ2020}. The boulders themselves
were heavily cracked (type B material), and most of them disintegrated into dust in the later stages of the atmospheric entry.
Only a small percentage of the original mass reached the ground as meteorites, mostly small ones. Four meteorites
larger than 100 g could be expected at most.

Ribbeck was not a particularly large meteorite fall, and yet, the meteoroid was discovered in space. This indicates that as the efficiency of asteroid discoveries increases and the bolide network expands, more such events can be expected. We have shown that valuable and accurate results can be obtained from bolide observations, provided the correct procedures are used. However, in terms of bolide observations, this means that the data acquired must be of sufficient quality and complexity.  This requires multi-instrument observations that include radiometers and spectroscopy in addition to standard camera systems. Only observations like this can provide the most complete description of not only the atmospheric trajectories and heliocentric orbits, but also of the physical properties and composition of meteoroids.

\begin{acknowledgements}
       We thank Radmila Bro\v{z}kov\'a from the Czech Hydrometeorogical Institute for providing us the 
       atmospheric wind model data and operators Sirko Molau (AMS16 Ketz\"{u}r) and Andre Kn\"{o}fel (AMS22 Lindenberg) for providing video recordings of the AllSky7 network from their sites. 
       We are grateful to Peter Brown for his quick and helpful review.
      This work was supported by grant no.\ 24-10143S from the Czech Science
      Foundation. The institutional research plan is RVO:67985815.
\end{acknowledgements}

%

\begin{thebibliography}{}

\bibitem[Borovička(1990)]{Bor90}
Borovička J. \ 1990,
\bac, 41, 391

\bibitem[Borovi{\v{c}}ka \& Berezhnoy(2016)]{Benesov-molecules}
Borovi{\v{c}}ka, J., \& Berezhnoy, A. A.\ 2016,
\icarus, 278, 248

\bibitem[Borovi{\v{c}}ka et al.(2019)]{IMC2018}
Borovi{\v{c}}ka, J., Spurn{\'y}, P., \& Shrben{\'y}, L.\ 2019,
in Proc. Int. Meteor Conf., Pezinok-Modra, Slovakia, 30 August -- 2 September 2018 
(R. Rudawska et al., eds.), p. 28

\bibitem[Borovi{\v{c}}ka et al.(2020)]{AJ2020}
Borovi{\v{c}}ka, J., Spurn{\'y}, P., \& Shrben{\'y}, L.\ 2020,
\aj, 160:42

\bibitem[Borovi{\v{c}}ka et al.(2022)]{Bor22}
Borovi{\v{c}}ka, J., Spurn{\'y}, P., Shrben{\'y}, L., et al.\ 2022,
\aap, 667:A157

\bibitem[Britt \& Consolmagno(2003)]{Britt}
Britt, T. D. \& Consolmagno, G. J.\ 2003, 
Meteorit. Planet. Sci., 38, 1161

\bibitem[Ceplecha(1987)]{Cep87}
Ceplecha, Z.\ 1987, 
\bac, 38, 222

\bibitem[Clark et al.(2004)]{Clark}
Clark, B. E., Bus, S. J., Rivkin, A. S. et al.\ 2004, 
\jgr, 109, E02001

\bibitem[Clark and Wiegert(2011)]{Clark11}
Clark, D. L., \&  Wiegert, P. A.\ 2011, 
Meteorit. Planet. Sci., 46, 1217

\bibitem[Colas et al.(2020)]{FRIPON} 
Colas, F., Zanda, B., Bouley, S., et al.\ 2020, 
\aap, 644, A53

\bibitem[Dibb et al.(2022)]{Dibb}
Dibb, S. D., Bell III, J. F. \& Garvie, L. A. J.\ 2022, 
Meteorit. Planet. Sci., 57, 1570

\bibitem[Gattacceca et al.(2022)]{Kindberg}
Gattacceca, J., McCubbin, F. M., Grossman, J. et al. \ 2022, 
Meteoritical Bulletin No.\ 110, 
Meteorit. Planet. Sci., 57, 2102

\bibitem[Hankey et al.(2020)]{MH20}
Hankey, M., Perlerin, V. \& Meisel, D.\ 2020, 
Planet. Space Sci, 190, 105005

\bibitem[Jenniskens et al.(2009)]{Jen09}
Jenniskens, P., Shaddad, M. H., Numan, D. et al. \ 2009,
Nature, 458, 7237, 485

\bibitem[Jenniskens et al.(2021)]{Jen21}
Jenniskens, P.,  Gabadirwe, M., et al. \ 2021,
Meteorit. Planet. Sci., 56, 844

\bibitem[Johnston \& Stern(2024)]{Johnston}
Johnston, C. O., \& Stern, E. C.\ 2024, 
\icarus, 408, 115807

\bibitem[Koschny \& Borovička(2017)]{definitions}
Koschny, D., \& Borovicka, J.\ 2017, 
WGN, J. IMO, 45, 91


\bibitem[M. P. C. Staff(2024)]{MPEC}
M. P. C. Staff 2024, 
Minor Planet Electronic Circular, 2024-B76

\bibitem[Meteoritical Bulletin(2024)]{MetBull}
Meteoritical Bulletin Database entry 2024, to be published in Meteorit. Planet. Sci., 
https://www.lpi.usra.edu/meteor/metbull.php?code=81447  

\bibitem[Spurn{\'y} et al.(2014)]{Benesov}
Spurn{\'y}, P., Haloda, J., Borovi{\v c}ka, J., et al. 2014,
\aap, 570:A39

\bibitem[Spurn{\'y} et al.(2017)]{Spu17}
Spurný, P., Borovička, J., Mucke, H., Svoreň, J.  2017,
\aap, 605:A68

\bibitem[Towner et al.(2022)]{Towner}
Towner, M. C., Jansen-Sturgeon, T., Cupák, M., et al.\ 2022,
Planet. Sci. J., 3:44

\bibitem[Vida et al.(2023)]{VidaACM}
Vida, D., Egal, A., Brown, P. G., et al.\ 2023,
Asteroids, Comets, Meteors Conference 2023 (LPI Contrib. No. 2851),
abstract \# 2049

\end{thebibliography}
%

\end{document}